\documentclass[twoside,twocolumn,9pt]{article}
\usepackage{extsizes}
\usepackage[super,sort&compress,comma]{natbib}
\usepackage[version=3]{mhchem}
\usepackage[left=1.5cm, right=1.5cm, top=1.785cm, bottom=2.0cm]{geometry}
\usepackage{balance}
\usepackage{mathptmx}
\usepackage{sectsty}
\usepackage{graphicx}
\usepackage{lastpage}
\usepackage[format=plain,justification=justified,singlelinecheck=false,font={stretch=1.125,small,sf},labelfont=bf,labelsep=space]{caption}
\usepackage{float}
\usepackage{fancyhdr}
\usepackage{fnpos}
\usepackage[english]{babel}
\addto{\captionsenglish}{%
  
}
\usepackage{array}
\usepackage{droidsans}
\usepackage{charter}
\usepackage[T1]{fontenc}
\usepackage[usenames,dvipsnames]{xcolor}
\usepackage{setspace}
\usepackage[compact]{titlesec}
\usepackage{hyperref}

\usepackage{physics}
\usepackage{wasysym}
\usepackage{xcolor}
\usepackage{amsmath}
\usepackage{amssymb}

\definecolor{cream}{RGB}{222,217,201}

\begin{document}

\pagestyle{fancy}
\thispagestyle{plain}
\fancypagestyle{plain}{
    \renewcommand{\headrulewidth}{0pt}
}

\makeFNbottom
\makeatletter
\renewcommand\LARGE{\@setfontsize\LARGE{15pt}{17}}
\renewcommand\Large{\@setfontsize\Large{12pt}{14}}
\renewcommand\large{\@setfontsize\large{10pt}{12}}
\renewcommand\footnotesize{\@setfontsize\footnotesize{7pt}{10}}
\makeatother

\renewcommand{\thefootnote}{\fnsymbol{footnote}}
\renewcommand\footnoterule{\vspace*{1pt}%
    \color{cream}\hrule width 3.5in height 0.4pt \color{black}\vspace*{5pt}}
\setcounter{secnumdepth}{5}

\makeatletter
\renewcommand\@biblabel[1]{#1}
\renewcommand\@makefntext[1]%
{\noindent\makebox[0pt][r]{\@thefnmark\,}#1}
\makeatother
\renewcommand{\figurename}{\small{Fig.}~}
\sectionfont{\sffamily\Large}
\subsectionfont{\normalsize}
\subsubsectionfont{\bf}
\setstretch{1.125} 
\setlength{\skip\footins}{0.8cm}
\setlength{\footnotesep}{0.25cm}
\setlength{\jot}{10pt}
\titlespacing*{\section}{0pt}{4pt}{4pt}
\titlespacing*{\subsection}{0pt}{15pt}{1pt}
\fancyfoot{}
\fancyfoot[LO,RE]{\vspace{-7.1pt}\includegraphics[height=9pt]{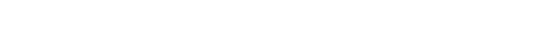}}
\fancyfoot[CO]{\vspace{-7.1pt}\hspace{13.2cm}\includegraphics{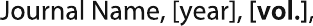}}
\fancyfoot[CE]{\vspace{-7.2pt}\hspace{-14.2cm}\includegraphics{head_foot/RF}}
\fancyfoot[RO]{\footnotesize{\sffamily{1--\pageref{LastPage} ~\textbar  \hspace{2pt}\thepage}}}
\fancyfoot[LE]{\footnotesize{\sffamily{\thepage~\textbar\hspace{3.45cm} 1--\pageref{LastPage}}}}
\fancyhead{}
\renewcommand{\headrulewidth}{0pt}
\renewcommand{\footrulewidth}{0pt}
\setlength{\arrayrulewidth}{1pt}
\setlength{\columnsep}{6.5mm}
\setlength\bibsep{1pt}

\makeatletter
\newlength{\figrulesep}
\setlength{\figrulesep}{0.5\textfloatsep}

\newcommand{\topfigrule}{\vspace*{-1pt}%
    \noindent{\color{cream}\rule[-\figrulesep]{\columnwidth}{1.5pt}} }

\newcommand{\botfigrule}{\vspace*{-2pt}%
    \noindent{\color{cream}\rule[\figrulesep]{\columnwidth}{1.5pt}} }

\newcommand{\dblfigrule}{\vspace*{-1pt}%
    \noindent{\color{cream}\rule[-\figrulesep]{\textwidth}{1.5pt}} }

\makeatother

\twocolumn[
    \begin{@twocolumnfalse}
        \vspace{1em}
        \sffamily
        \begin{tabular}{m{4.5cm} p{13.5cm} }
            \includegraphics{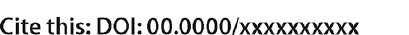}   & \noindent\LARGE{\textbf{Deformation and orientational order of chiral membranes with  free edges}}\\
            \vspace{0.3cm} & \vspace{0.3cm}\\
            & \noindent\large{Lijie Ding,$^{\ast}$\textit{$^{a}$} Robert A. Pelcovits,\textit{$^{ab}$} and Thomas R. Powers\textit{$^{abcd}$}}\\
            \includegraphics{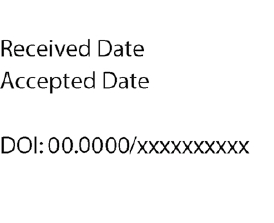} & \noindent\normalsize{Motivated by experiments on colloidal membranes composed of chiral rod-like viruses, we use Monte Carlo methods to determine the phase diagram for the liquid crystalline order of the rods and the membrane shape. We generalize the Lebwohl-Lasher model for a nematic with a chiral coupling to a curved surface with edge tension and a resistance to bending, and include an energy cost for tilting of the rods relative to the local membrane normal. The membrane is represented by a triangular mesh of hard beads joined by bonds, where each bead is decorated by a director. The beads can move, the bonds can reconnect and the directors can rotate at each Monte Carlo step. When the cost of tilt is small, the membrane tends to be flat, with the rods only twisting near the edge for low chiral coupling, and remaining parallel to the normal in the interior of the membrane. At high chiral coupling, the rods twist everywhere, forming a cholesteric state. When the cost of tilt is large, the emergence of the cholesteric state at high values of the chiral coupling is accompanied by the bending of the membrane into a saddle shape. Increasing the edge tension tends to flatten the membrane. These results illustrate the geometric frustration arising from the inability of a surface normal to have twist.
            }\\
        \end{tabular}
    \end{@twocolumnfalse} \vspace{0.6cm}
]

\renewcommand*\rmdefault{bch}\normalfont\upshape
\rmfamily
\section*{}
\vspace{-1cm}

\footnotetext{\textit{$^{\ast}$~Email: Lijie\_Ding@brown.edu}}
\footnotetext{\textit{$^{a}$~Department of Physics, Brown University, Providence, RI 02912, USA.}}
\footnotetext{\textit{$^{b}$~Brown Theoretical Physics Center, Brown University, Providence, RI 02912, USA.}}
\footnotetext{\textit{$^{c}$~School of Engineering, Brown University, Providence, RI 02912, USA.}}
\footnotetext{\textit{$^{d}$~Center for Fluid Mechanics, Brown University, Providence, RI 02912, USA.}}



\section{Introduction}
    Chirality arises at various length scales of soft matter systems,\cite{amabilino2009chirality,bahr2001chirality,wagniere2007chirality} and can play a central role in determining the internal order of the system, as in the transitions between the various phases of cholesteric liquid crystals, namely the isotropic phase, the blue phase and the helical phase.\cite{wright1989crystalline}

    Fluid membranes are also ubiquitous in soft matter systems and exhibit various topologies and shapes. For a closed membrane, vesicles, pears, discocytes, stomatocytes and toroids\cite{seifert1997configurations} are all possible shapes. As for membranes with open edges, they can also form various shapes including disks, scallops, ribbons and starfish.\cite{gibaud2017filamentous} Membranes made of rod-like chiral particles that tend to align with the surface normal experience geometric frustration: it is impossible for the rods to follow the preferred cholesteric twist and remain normal the membrane surface. This frustration is analogous to the frustration experienced by cholesteric phases confined between two parallel plates with hometropic boundary conditions, or subject to external electric or magnetic fields.\cite{KamienSelinger2001,OswaldPieranski2005,duzgun2018comparing} In this paper, we explore the shapes and liquid crystalline phases displayed by a model colloidal membrane system with chiral rod-like constituents.

    A two dimensional colloidal membrane with free edges and composed of chiral rod-like viruses\cite{gibaud2012reconfigurable} is an example where both chirality and membrane deformability are key components of the system. The interplay of chirality and deformability leads to changes in the membrane's mechanical properties\cite{balchunas2020force} and shapes,\cite{jia2017chiral} including the formation of three dimensional structures.\cite{sharma2020,Robaszewski2020} Although many theoretical models have been developed for the chiral membranes with free edges, including phenomenological theories,\cite{tu2003lipid,kaplan2010theory,tu2013theory} entropically-motivated theories\cite{kang2016entropic,gibaud2017achiral} and effective energy theories,\cite{jia2017chiral,balchunas2020force} theoretical analyses usually require an \textit{a priori} assumption of the shape of the membrane. Numerical simulations should be able to avoid this assumption and predict a shape phase diagram. However, computational studies of two-dimensional colloidal membranes including the depletion effect have been limited primarily to hard body simulations of flat membranes.\cite{yang2012self,xie2016probing} Accounting for these depletion effects for curved colloidal membranes is computationally costly. The only Monte Carlo simulations of membranes with orientational order and curved shapes have been of lipid bilayer vesicles with in-plane orientational order.\cite{koibuchi2008possible,ramakrishnan2010monte,nguyen2013nematic,sreeja2015monte} These and other theoretical studies\cite{LubenskyMacKintosh1993,SelingerSchnur1993} typically consider a \textit{constant} angle of tilt between the nematic director and the surface normal. Colloidal membranes, on the other hand, exhibit smectic order with variable director tilt, with zero tilt in most of the membrane interior and nonzero tilt near the membrane edges or at interior $\pi$ walls\cite{piwalls} or the boundaries of rafts of short virus rods in a background of long virus rods.\cite{rafts} In our previous work,\cite{ding2020shapes} we took a step towards developing a more general computational approach that allows for director tilt and arbitrary membrane shapes. We developed a Monte Carlo simulation scheme for a chiral membrane with free edges using a discretized effective energy. In this model we did not treat the liquid crystal director degrees of freedom directly; rather, we employed an effective energy\cite{jia2017chiral} where chirality and Frank elasticity are modeled by suitable edge geometric quantities. Such an approach is reasonable if the chiral twist is confined to the edge as it is in large flat membranes but not more generally. A more comprehensive model should include the full liquid crystalline degrees of freedom on the entire membrane and study the coupling between the orientational order and membrane surface shape.

    In this paper, we introduce such a model where the shape of the membrane is modeled by a triangular mesh as in our previous work, and the orientational degrees of freedom are introduced by decorating each vertex of the mesh with a unit-vector director. The energy for the membrane is inspired by a phenomenological model:\cite{kaplan2010theory} we use the discretized Canham-Helfrich bending energy and a line tension energy for the membrane shape, the Lebwohl-Lasher interaction for the directors, a pseudoscalar proportional to the twist of the neighboring directors, and finally, a tilt coupling energy which favors the alignment of director and the local surface normal. For the purposes of the present study, the bending moduli and line tension are tuned such that the membranes have the topology of a disk rather than that of a closed vesicle or the shape of a branched polymer. We start by investigating the director field, and find three phases: isotropic, smectic-A and cholesteric, depending on chirality and the strength of the Lebwohl-Lasher coupling. Detailed studies are then carried out for the smectic-A and cholesteric phases. We find that in the cholesteric phase the membrane does not remain flat but bends into a saddle-like shape instead. We develop a simple model to understand this phenomenon.

    The rest of this paper is organized as follows. In Section~\ref{sec:model_and_method}, we define our discrete model and explain the Monte Carlo method we use for the simulations. We present the results of our simulations in Section~\ref{sec:results}. Finally, we conclude our paper in Section~\ref{sec:conclusion}.

\section{Model and Method}
\label{sec:model_and_method}
\subsection{Membrane with director field}
\begin{figure}[tbh]
        \includegraphics[width=\linewidth]{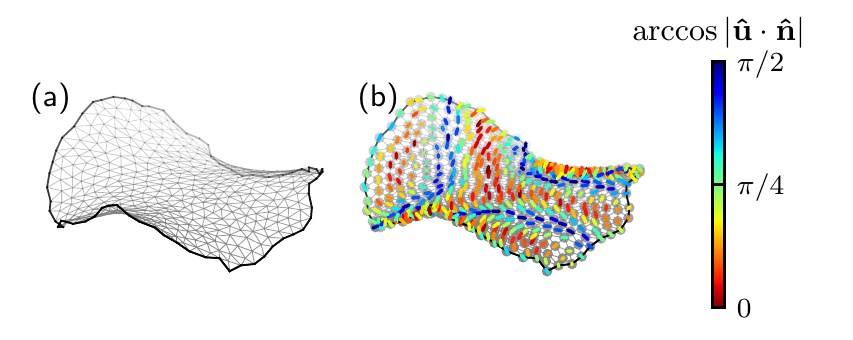}
        \caption{\label{fig:config_demo} A configuration of the discretized membrane with directors.
        (a) The membrane shape is modeled by a triangular mesh of bonds with hard sphere beads (not shown) located at the vertices of the mesh.
        (b) Here the directors $\vu{u}_i$ on the beads are shown; the color of the director is determined by the tilt  relative to the local surface normal $\vu{n}_i$, as indicated by the color bar on the right.}
    \end{figure}

    We model the membrane using a bond-and-bead triangular mesh $\mathcal{M}$ for self-avoiding membranes,\cite{gompper1997network} with a bead located at each vertex $i$ of the mesh and decorated with a unit-vector director $\vu{u}_i$ (Fig.~\ref{fig:config_demo}). The beads are hard spheres of diameter $\sigma_0$ connected by bonds of maximum length $l_0$. The directors are free to rotate in three-dimensional space.

    The total energy $E$ of the membrane is a sum of a surface energy $E_s$, dependent only on the shape represented by the triangular mesh, and a liquid-crystalline energy $E_{lc}$ arising from the director field and its coupling with the membrane shape. The surface energy $E_s$ is a sum of the discretized Canham-Helfrich bending energy\cite{ding2020shapes,canham1970minimum,helfrich1973elastic} and a membrane edge energy:

    \begin{equation}
        \label{eq:E_s}
        E_s = \frac{\kappa}{2}\sum_{i\in\mathring{\mathcal{M}}} (2H_i)^2\sigma_i +\lambda \sum_{i\in \partial \mathcal{M}}\dd{s}_i,
    \end{equation}

    where $\kappa$ is the membrane bending modulus, and $H_i$ and $\sigma_i$ are the mean curvature and the area of the cell on the virtual dual lattice at bead $i$, respectively. Complete expressions for these quantities can be found elsewhere.\cite{gompper1997network,ding2020shapes,espriu1987triangulated} The modulus $\lambda$ is the line tension and $\dd{s}_i$ is the differential edge length at bead $i$.  The summation in the first term on the right hand side of eqn~(\ref{eq:E_s}) is over all interior beads $\mathring{\mathcal{M}}$ of the mesh, while the summation in the last term is over the edge $\partial\mathcal{M}$ of the mesh.  We assume that the Gaussian curvature modulus of the Canham-Helrich model is zero and limit our study to the interplay of the chirality and the coupling of liquid crystalline order of the rods with the shape.

    The liquid-crystalline energy $E_{lc}$ is given by the sum of three contributions: $E_{lc}=E_{LL}+E_c+E_t$, where $E_{LL}$ is a Lebwohl-Lasher interaction\cite{lebwohl1972nematic} that favors the alignment of neighboring directors, $E_c$ is a discrete chiral energy favoring  twist of neighboring directors and $E_t$ is an effective tilt energy favoring alignment of the director and local surface normal. The Lebwohl-Lasher energy $E_{LL}$ is given by

    \begin{equation}
        \label{eq:E_LL}
        E_{LL} = -\epsilon_{LL} \sum_{(i,j)\in\mathcal{B}} \left[\frac{3}{2}(\vu{u}_i\cdot\vu{u}_j)^2-\frac{1}{2}\right],
    \end{equation}

    where $\epsilon_{LL}$ denotes the interaction constant and the summation is over all bonds $\mathcal{B}$ joining neighboring directors in the triangular mesh. The effective chiral energy $E_c$ is constructed from a chiral pseudoscalar used in the chiral Lebwohl-Lasher model\cite{memmer2000computer} and is given by

    \begin{equation}
        \label{eq:E_c}
        E_c = -\epsilon_{c} \sum_{(i,j)\in\mathcal{B}}(\vu{u}_i\cross\vu{u}_j)\cdot\vu{r}_{ij}(\vu{u}_i\cdot\vu{u}_j),
    \end{equation}

    where $\epsilon{c}$ is the chiral interaction parameter and $\vu{r}_{ij}$ is the unit vector parallel to the bond $(i,j)$. The Lebwohl-Lasher and chiral interactions lead to a preferred angle of twist for a single pair of directors of $\arctan(2k_c/3)/2$, where $k_c=\epsilon_c/\epsilon_{LL}$. Finally, the tilt energy $E_t$ is given by

    \begin{equation}
        \label{eq:E_t}
        E_t = \frac{1}{2}C\sum_{i\in\mathcal{M}} \left[1-(\vu{u}_i\cdot \vu{n}_i)^2\right],
    \end{equation}

    where $C$ is the tilt coupling constant and $\vu{n}_i$ is the surface normal at bead $i$. When $C>0$, as in our study, alignment between the director and local surface normal is favored. The total energy $E$ of the membrane is then given by the sum of eqn~(\ref{eq:E_s})-(\ref{eq:E_t}).

\subsection{Monte Carlo method}
    To sample the configuration space of the model, the beads, bonds and directors on the triangular mesh are all subject to updates. The bead and bond updates follow the same procedure described in our previous paper\cite{ding2020shapes} and other studies.\cite{boal1992topology,gompper1997network} To update the director field, we follow the algorithm described by Baker et al.,\cite{barker1969structure} which consists of the following steps: (i) choose a director at random, (ii) choose a rotation axis at random from the global coordinate axes $(x,y,z)$ with equal probability, (iii) rotate the director about the chosen axis by an angle selected from the uniform probability distribution $[-\delta\phi,\delta\phi]$.

    In our simulation, $6\times10^3$ MC steps were performed. For a system of $N$ beads, each step is composed of $N/t^2$ attempts to move a bead chosen at random, $2N/t^2$ attempts to flip a bond chosen at random and $\sqrt{N}/t^2$ attempts to shrink or extend the edge of the membrane. Here $t$ gives the range of the bead position updates: a bead chosen at random is moved with uniform probability to a new position in a cube of side $2t$ centered at the original position. The parameter $t$ is set to $0.1$, with all lengths measured in units of the bead diameter $\sigma_0$. The initial membrane configuration is a circular disk in the $xy$ plane with all directors pointing the $z$ direction. We first equilibrate the system for $2\times10^3$ MC steps, then record the data for every subsequent MC step. All observables are measured for $4\times10^3$ MC steps. The uncertainty in the observables is estimated using Sokal's method\cite{sokal1997monte}. The director rotation parameter is $\delta\phi=0.5$ and, to ensure the fluidity of the membrane,\cite{gompper2000melting} the maximum bond length is set to $l_0=1.73$. We choose the bending modulus $\kappa=100$ (all energies are measured in units of $k_BT)$ and the number of beads $N=439$. These latter choices guarantee that we avoid transitions to either a branched polymer shape or closed vesicle.\cite{ding2020shapes,boal1992topology}

\section{Results}
\label{sec:results}
\subsection{Director field patterns}
\begin{figure}[tbh]
        \includegraphics[width=\linewidth]{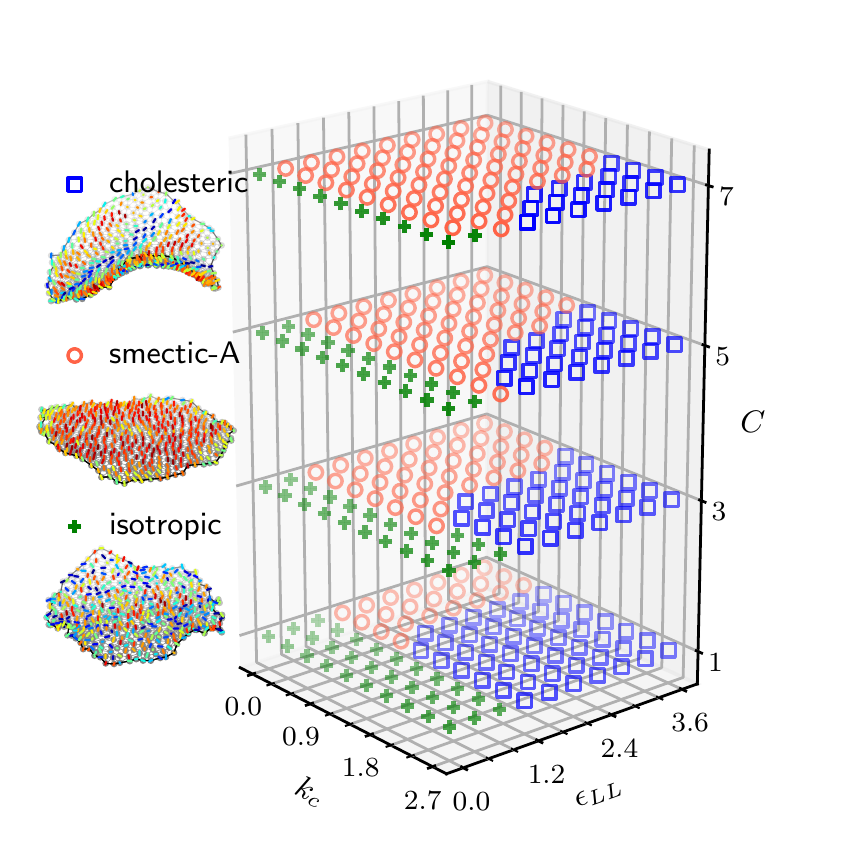}
        \caption{\label{fig:phase_diagram} Phase diagram for the director field on a membrane with $N=439$, $\kappa=100$ and $\lambda=6$. Isotropic, smectic-A and cholesteric phases are represented by {\color{green}+}, {\color{red}$\bigcirc$} and {\color{blue}$\square$}, respectively. Sample configurations on the left side, from bottom to top, have $(\epsilon_{LL},C,k_c)$ equal to $(0,0,0)$, $(3.2,3.0,0.9)$ and $(3.2,3.0,1.5)$, for isotropic, smectic-A and cholesteric, respectively. The color of the directors is a measure of the tilt angle, i.e. the angle between the director and the local normal to the membrane. See Fig.~\ref{fig:config_demo} for the color bar.}
    \end{figure}

    We first explore the phase diagram associated with the orientational order of the director field. We find that director field can form three different patterns---isotropic, smectic-A and cholesteric---as we vary the liquid-crystalline energy parameters, $\epsilon_{LL}$, $k_c$ and $C$. As shown in Fig.~\ref{fig:phase_diagram}, an isotropic, (orientationally disordered) phase appears, not surprisingly, when the Lebwohl-Lasher interaction $\epsilon_{LL}$ is relatively weak. The critical value of $\epsilon_{LL}$ below which the isotropic phase forms decreases as the tilt coupling constant $C$ increases, which is due to the flatness of the membrane shape and the alignment between the director and the membrane surface normal $C$ encourages. Above this critical value of $\epsilon_{LL}$ and for sufficiently small twist constant $k_c$, a chiral smectic-A phase forms with the directors aligned in the interior of the membrane, while the twist of the directors is expelled to the edge of the membrane, as first predicted by de Gennes.\cite{de1972analogy} As $k_c$ increases, the twist penetrates into the interior of the membrane leading to the formation of $\pi$ walls. We consider this state to be a cholesteric phase, although the $\pi$ walls are not exactly parallel. In the phase diagram, we defined the isotropic phase to be the region with $\left<\frac{3}{2}(\vu{u}_i\cdot\vu{u}_j)^2-\frac{1}{2}\right>_{(i,j)}<0.5$, where $\left<\dots\right>_{(i,j)}$ denotes the average over all bonds. We determined the boundary between the smectic-A phase and the cholesteric phase using $\left<(\vu{u}_i\cross\vu{u}_j)\cdot\vu{r}_{ij}(\vu{u}_i\cdot\vu{u}_j)\right>_{(i,j)}=0.1$.

\begin{figure}[tbh]
        \includegraphics[width=\linewidth]{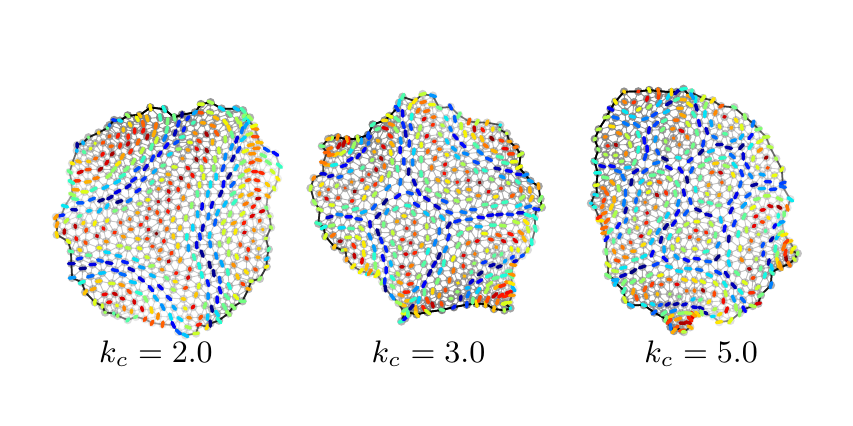}
        \caption{\label{fig:config_meron} Top view of the configurations of a membrane with $N=439$, $\kappa=100$, $\lambda=6$ and $\epsilon_{LL}=C=3$, for $k_c=2$, $3$ and $5$ (from left to right). The three configurations shown are the cholesteric phase ($k_c=2$), the formation of a meron lattice ($k_c=3$) and a meron lattice ($k_c=5$). The color of the directors is a measure of the tilt angle of the director with respect to the local normal to the membrane; see Fig.~\ref{fig:config_demo} for the color bar.}
    \end{figure}

    A similar phase diagram was found by Duzgun et al.\cite{duzgun2018comparing} in their theoretical and computational studies of a flat two-dimensional model of chiral liquid crystals with no boundary. Our model is distinguished from that of Duzgun et al. by its finite size and edge energy and, more importantly, by the deformability of the membrane. In the model studied by Duzgun et al., our tilt coupling is analogous to an interaction with an electric field. For a positive dielectric anisotropy, the tilt and electric field interactions are mathematically identical. The case of negative dielectric anisotropy is equivalent to a tendency for the directors to lie in the local tangent plane of the membrane. This case has no counterpart in the virus membranes of interest to us, and is therefore disregarded in the present study. Similar to our results, Duzgun et al. found isotropic, vertical nematic (our Smectic-A) and cholesteric phases. They also found meron phases and metastable skyrmion phases\cite{Lin2015}. The meron phase has regions of double twist separated by $\pi$ walls that meet in three-fold junctions at a disclination. The skyrmion phase has double twist with no singularities. We also find a stable meron phase (Fig.~\ref{fig:config_meron}) for values of $k_c$ larger than those shown in Fig.~\ref{fig:phase_diagram}. Duzgun et al. studied the cholesteric to meron lattice phase transition by calculating the free energy of the two phases. Because we have found that the cholesteric to meron lattice transition is not accompanied by a discernible membrane shape change, we do not consider the cholesteric to meron lattice transition in detail.

\subsection{Twist penetration}
\begin{figure}[htb]
        \includegraphics[width=\linewidth]{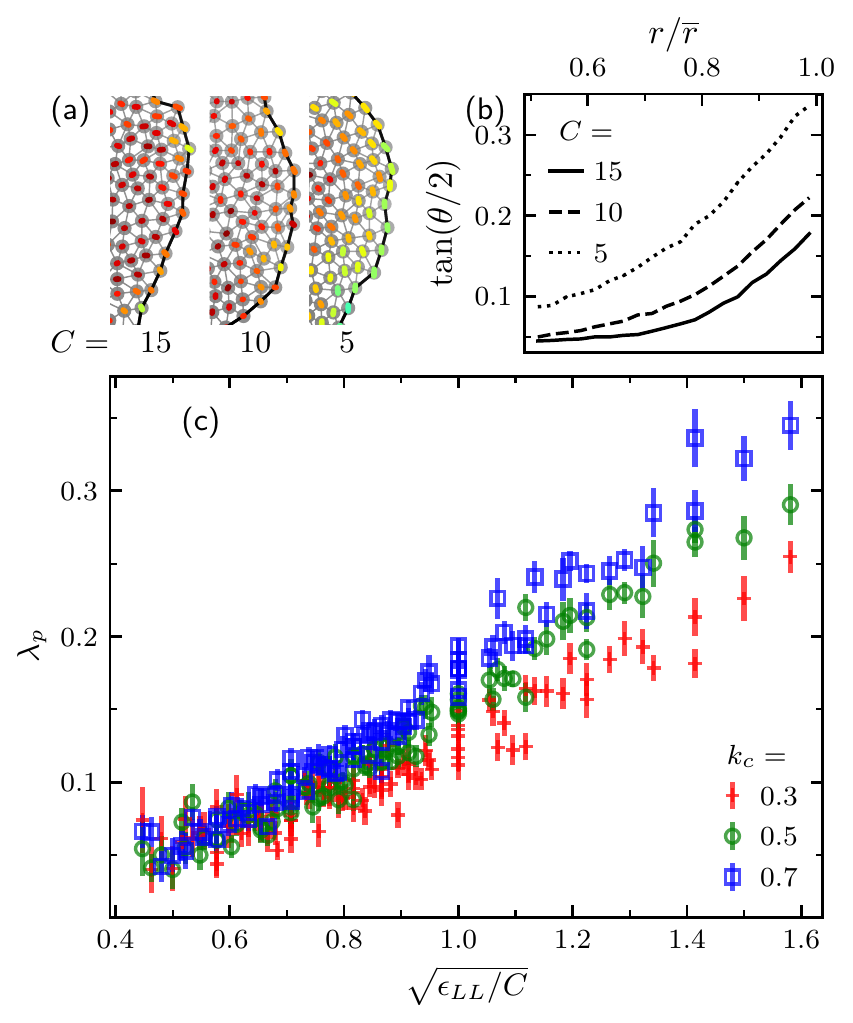}
        \caption{\label{fig:penetration_depth} Penetration of the director twist in the smectic-A phase of membranes with $N=439$, $\kappa=100$ and $\lambda=6$.
        (a) Bead, bond and director configurations near the membrane edge for for three different values of $C$ with $\epsilon_{LL}=10$ and $k_c=0.7$. The color scheme for the tilt angle is the same as in Fig.~\ref{fig:config_demo}. Note the greater penetration depth for smaller $C$.
        (b) A plot of $\tan(\theta/2)$, ($\theta$ is the director tilt angle) versus distance from the center of mass of membrane, normalized to the average distance $\overline{r}$ from the center, for three values of $C$. Here, $\epsilon_{LL}=10$ and $k_c=0.7$.
        (c) Penetration depth calculated by fitting to the expression $\tan(\theta/2)=\tan(\theta_0/2)\exp[(r/\overline{r}-1)/\lambda_p]+\textit{const.}$ as the ratio $\epsilon_{LL}/C$ is varied. Data is shown for $k_c=0.3(+)$, $0.5(\circ)$ and $0.7(\square$) with corresponding coefficient of determination values $R^2$ of $0.90$, $0.96$ and $0.96$, respectively,}
    \end{figure}

    In a chiral smectic-A membrane the director twist is expelled to the edge and, as shown by de Gennes,\cite{de1972analogy} the twist penetration depth $\lambda_p$ is proportional to the square root of the ratio of the twist Frank elastic constant (in the present case, the Lebwohl-Lasher interaction $\epsilon_{LL}$) and the tilt modulus $C$. Fig.~\ref{fig:penetration_depth}(a) shows a view of a portion of the membrane edge for three values of $C$. As $C$ decreases, the twist penetrates further into the membrane bulk, as expected from de Gennes' prediction. By measuring the tilt angle $\theta=\arccos(|\vu{u}\cdot\vu{n}|)$ between the director and local surface normal at each bead, and the distance $r$ to the center-of-mass of the membrane, we can quantify the decay of tilt from the edge to the bulk [Fig.~\ref{fig:penetration_depth}(b)]. In Fig.~\ref{fig:penetration_depth}(b) we normalize the radius $r$ with $\overline{r}$, the average distance from the center-of-mass to the membrane perimeter. For sufficiently large membranes it is expected on theoretical and experimental grounds \cite{pelcovits2009twist,barry2009direct} that $\tan(\theta/2)$ grows exponentially near the edge: $\tan(\theta/2)=\tan(\theta_0/2)\exp[(r/\overline{r}-1)/\lambda_p]+\textit{const}$, where $\theta_0$ is the value of the tilt angle at the membrane edge. The fit of our simulation data to this latter expression is shown in Fig.~\ref{fig:penetration_depth}(c) for three different values of $k_c$. There is good agreement with de Gennes' prediction that $\lambda_p\propto\sqrt{\epsilon_{LL}/C}$.

\subsection{Smectic-A to cholesteric transition}
\begin{figure}[htb]
        \includegraphics[width=\linewidth]{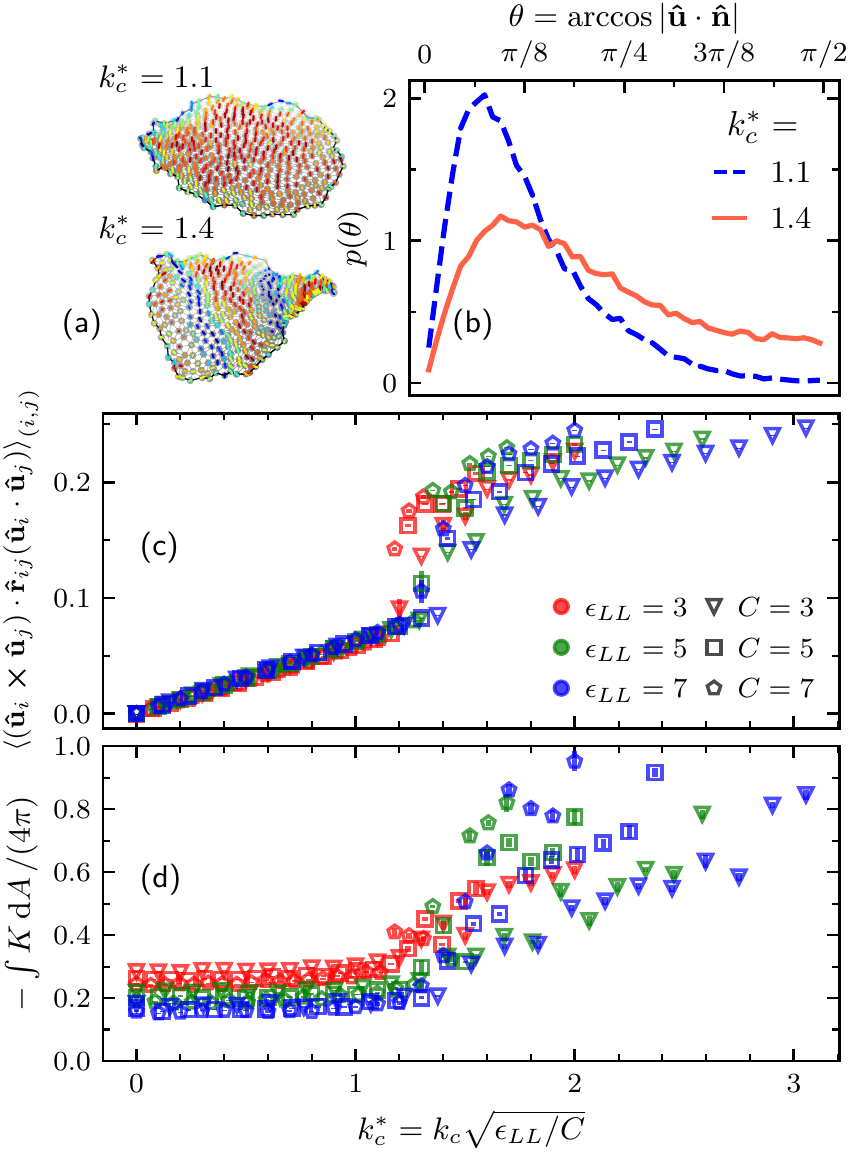}
        \caption{\label{fig:O_kc_reduced} Orientational and geometric properties of the membrane in the  smectic-A and cholesteric phases with $N=439$, $\kappa=100$ and $\lambda=6$.
        (a) Typical configurations of the membrane in the smectic-A (top) and cholesteric (bottom) phases for $\epsilon_{LL}=C=3$ for $k_c=1.1$ and $1.4$ with director field colored according to the tilt angle same as in Fig.~\ref{fig:config_demo}.
        (b) Normalized distribution $p(\theta)$ of the tilt angle $\theta=\arccos{|\vu{u}\cdot\vu{n}|}$ on a membrane with $\epsilon_{LL}=C=3$ for $k_c=1.1$ (dashed line) and $k_c=1.4$ (solid line).
        (c) Average twist per bonded pair of beads versus reduced twist constant $k_c^*=k_c\sqrt{\epsilon_{LL}/C}$ for various combinations of $\epsilon_{LL}$ and $C$.
        (d) Integral of the \textit{negative} Gaussian curvature versus $k_c^*$. The color coding and symbol are the same as in (c). The integral has been normalized to the its value for a sphere.}
    \end{figure}

    We now examine the smectic-A to cholesteric transition of the director field. Fig.~\ref{fig:O_kc_reduced}(a) shows typical configurations of the membrane in the smectic-A and cholesteric phases. The normalized distribution $p(\theta)$ of the tilt of the director with respect to the local layer normal [see Fig.~\ref{fig:O_kc_reduced}(b)] clearly shows the difference in the orientational order of the two phases. The distribution in the smectic-A phase has a strong peak near $\theta\sim 0$. The deviation from $\theta=0$ is due to thermal fluctuations. The distribution in the cholesteric phase has a weaker peak for $\theta\sim0$ and a longer tail not reaching zero. These features are associated with the formation of $\pi$ walls (twist walls) where the directors rotate through $180^\circ$. While the Lebwohl-Lasher interaction $\epsilon_{LL}$ and the tilt modulus $C$ are the two main competing factors for the tilt of director field, the smectic-A to cholesteric transition is mainly driven by the twist constant $k_c$. This can be seen in Fig.~\ref{fig:O_kc_reduced}(c) where the average twist $\left<(\vu{u}_i\cross\vu{u}_j)\cdot\vu{r}_{ij}(\vu{u}_i\cdot\vu{u}_j)\right>_{(i,j)}$ between directors joined by bonds is plotted versus the reduced twist constant $k_c^*=k_c\sqrt{\epsilon_{LL}/C}$. In these expressions the average $\left<\dots\right>_{(i,j)}$ is taken over all bonds on the membrane, i.e., $(i,j)\in\mathcal{B}$. The average twist data collapses to a single line when the membrane is in the smectic-A phase. Furthermore, the value of $k_c^*$ at the transition is not very sensitive to the value of $\epsilon_{LL}$. The data ceases to collapse in the cholesteric phase. Of greater interest is that the membrane shape changes at the smectic-A to cholesteric transition as can be seen in Fig.~\ref{fig:O_kc_reduced}(d) where the integral of the negative Gaussian curvature is plotted as a function of $k_c^*$. As can be seen from Figs.~\ref{fig:O_kc_reduced}(a) and (d), the smectic-A to cholesteric transition is accompanied by a change in the shape of the membrane, in particular to a shape with negative Gaussian curvature. We computed the Gaussian curvature on the triangular mesh using methods found elsewhere.\cite{ding2020shapes,meyer2003discrete} We explore this shape change in the next section.

\subsection{Membrane shape change at the smectic-A to cholesteric transition}
\begin{figure}[htb]
        \includegraphics[width=\linewidth]{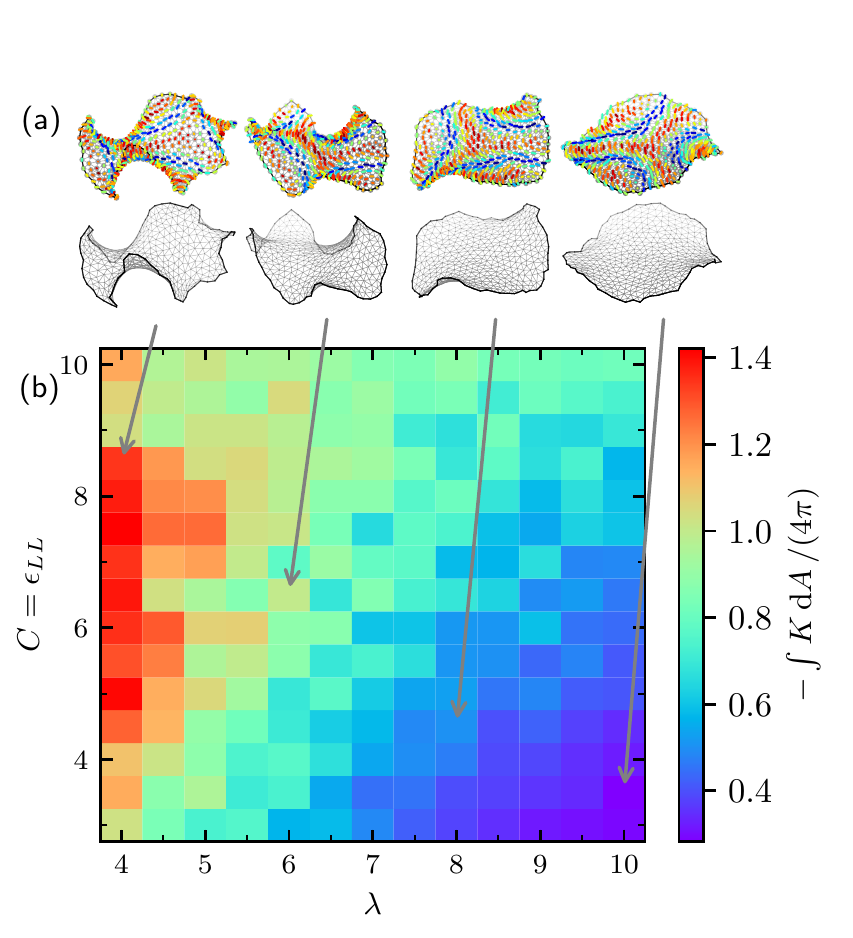}
        \caption{\label{fig:negK} Membrane shape changes accompanying the formation of the cholesteric phase. Here, $N=439$, $\kappa=100$, $k_c=2.0$ and $\epsilon_{LL}=C$. Thus, the value of the reduced twist constant $k_c^*=k_c\sqrt{\epsilon_{LL}/C}=2$ for all data shown in this figure.
        (a) Configuration of the membrane for values of $(\lambda,C=\epsilon_{LL})$ corresponding to points indicated with arrows in part (b). The top row shows the membrane with the director field colored according to the tilt angle (see Fig.\ref{fig:config_demo} for the color bar). The bottom row shows the same membrane with the directors removed.
        (b) Heat map of the integral of the normalized negative Gaussian curvature as a function of the line tension $\lambda$, Lebwohl-Lasher coupling $\epsilon_{LL}$ and tilt coupling $C$ with $\epsilon_{LL}=C$. The integral has been normalized to its value for a sphere.}
    \end{figure}

    From Fig.~\ref{fig:O_kc_reduced}(c) we see that the transition from smectic-A to cholesteric is controlled by the reduced twist constant $k_c^*=k_c\sqrt{\epsilon_{LL}/C}$. Thus, to explore the shape of the membrane in the cholesteric phase, we keep the value of $k_c^*$ fixed by setting $k_c=2$ and $\epsilon_{LL}/C=1$. In Fig.~\ref{fig:negK}(a) examples of membrane configurations in the cholesteric phase for $k_c^*=2$ are shown with and without the director field (for clarity). The arrows in the figure indicate the value of $\lambda$ and the common value of $C$ and $\epsilon_{LL}$ corresponding to the shape shown. Fig.~\ref{fig:negK}(b) shows the normalized integral of the negative Gaussian curvature $-\int K \dd{A}/(4\pi)$ for different values of the tilt coupling $C$ ($=\epsilon_{LL}$) and the line tension $\lambda$. As indicated by the color bar, the integral of the negative Gaussian curvature decreases from the upper left to the lower right of the plot, whereas ratio $\lambda/C$ increases as we move in the same direction.

    Our interpretation of Fig.~\ref{fig:negK} is as follows. When $\lambda/C$ is small, the membrane energy is dominated by the tilt interaction, which favors alignment between the surface normal and director. In the cholesteric phase, there is twist everywhere in the interior of the membrane, instead of only at the edge as in the smectic-A phase. Thus, the membrane surface tends to deform into a saddle shape to lower the tilt energy by making the normal vector align more closely with the directors over part of the membrane area. On the other hand, bending a flat disk at fixed area into to a saddle shape increases the perimeter, leading to an energy cost proportional to the line tension $\lambda$. Thus, increasing the line tension favors a disk shape. Although we study the case with $\epsilon_{LL}=C$, it is important to note that increasing $\epsilon_{LL}$ can have the \textit{opposite} effect of increasing $C$. For a fixed total number of beads, our mesh has more bonds if the shape is a disk than if the shape is saddle-like, since the disk configuration has fewer beads on the edge. Therefore, increasing the Lebwohl-Lasher parameter can lead to a preference for disks as it favors more bonds. Apparently, this tendency dominates at the lower values of $\lambda$ in Fig.~\ref{fig:negK}(b).

    \begin{figure}[htb]
        \includegraphics[width=\linewidth]{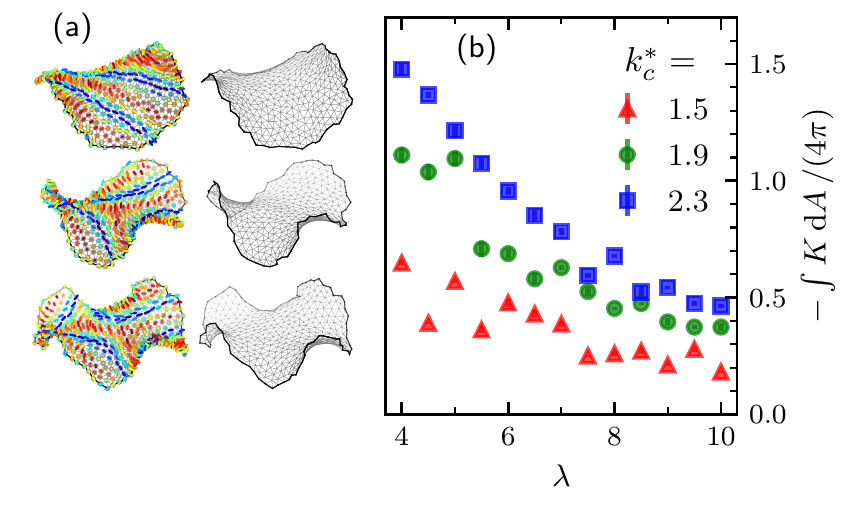}
        \caption{\label{fig:negK_lam} Membrane rippling associated with $\pi$ walls intersecting the membrane edge. Here $N=439$, $\kappa=100$ and $\epsilon_{LL}=C=5$.
        (a) Membrane configurations for $k_c=1.5$ (top), $1.9$ (middle) and $2.3$ (bottom), with $\lambda=6$. The left column shows the director fields colored by the local tilt angle as in Fig.~\ref{fig:config_demo}. The right column shows the same membrane but with directors removed.
        (b) The negative integrated Gaussian curvature decreases as line tension $\lambda$ increases, with a rate depending on $k_c^*$.}
    \end{figure}

    From Figs.~\ref{fig:negK} and \ref{fig:negK_lam} we see that the number of ripples is related to the number of $\pi$ walls intersecting the membrane edge. The top figure of Fig.~\ref{fig:negK_lam}(a) shows the membrane shape becomes saddle-like when there are two $\pi$ walls intersecting the edge. As $k_c^*$ increases, more $\pi$ walls intersect the membrane edge and the saddle shapes become more rippled.

\begin{figure}[htb]
        \includegraphics[width=\linewidth]{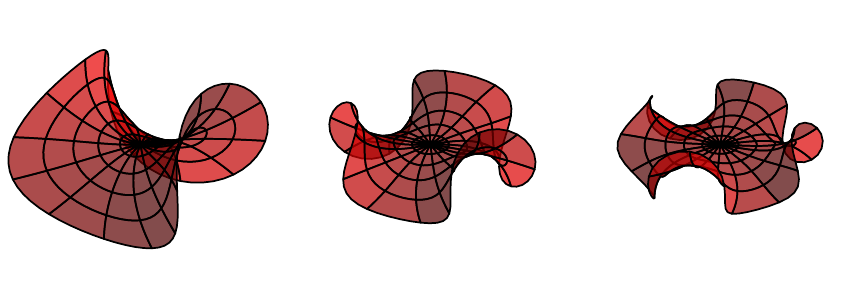}
        \caption{\label{fig:Enneper_shape} Enneper's surface with $r_1=1$ for $m=1$ (left), $m=2$ (middle) and $m=3$ (right). The wire frames are a guide to the eye.}
    \end{figure}

    Because we have chosen a large value of the membrane bending modulus $\kappa=100$ (motivated by experiments on colloidal membranes composed of rod-like viruses,\cite{gibaud2012reconfigurable} and by simulations of stiff membranes\cite{ramakrishnan2010monte}), the shapes we find are nearly minimal surfaces with $\int{(2H)^2\dd{A}}/(16\pi)=0.074\pm0.002$, allowing us the construct a simple model to better understand the shape changes associated with the smectic-A to cholesteric transition. We model the membrane as an Enneper surface which is a minimal surface (necessarily with negative Gaussian curvature) and resembles the structures we see in our simulations. The $m$th order Enneper surface of area $A$ is parameterized by the coordinates $(r,\phi)$ by:~\cite{enneper}

    \begin{equation}
        \label{eq:Enneper_surface}
        \begin{aligned}
            x/R & = r\cos(\phi)-\frac{r^{2m+1}}{2m+1}\cos[(2m+1)\phi]  \\
            y/R & = -r\sin(\phi)-\frac{r^{2m+1}}{2m+1}\sin[(2m+1)\phi] \\
            z/R & = \frac{2r^{m+1}}{m+1}\cos[(m+1)\phi]
        \end{aligned}
    \end{equation}

    where $r\in[0,r_1]$, $\phi\in[0,2\pi)$ and

    \begin{equation}
        \label{eq:Rr1}
        \frac{1}{R}=r_1\sqrt{\frac{\pi}{A}}\sqrt{1+\frac{2r_1^{2m}}{m+1}+\frac{r_1^{4m}}{2m+1}}
    \end{equation}

    is the normalization factor which keeps the area equal to $A$. The parameter $r_1$ controls the amplitude of the ripples, and $m$ controls the number of ripples. Examples of the Enneper surface for $m=1$, $2$ and $3$ are shown in Fig.~\ref{fig:Enneper_shape}.

    We embed the surface in a three-dimensional cholesteric phase. For simplicity we assume that the director field on the membrane $\vu{u}(x,y,z)$ is determined by a twist wavevector $\va{q}=q\vu{q}$ lying in the $xy$ plane:

    \begin{equation}
        \label{eq:3d_cholesteric}
        \vu{u}(x,y,z) = (\vu{z}\cross\vu{q})\sin(\va{q}\cdot\va{r})+\vu{z}\cos(\va{q}\cdot\va{r}).
    \end{equation}

    We now compare the energies of a flat disk and an Enneper surface, each having an embedded director field given by eqn~\ref{eq:3d_cholesteric}. Because we embed the membrane surface in a fixed director field, the liquid crystalline twist energy is the same for both shapes; thus, the relative energy of these two surfaces is determined by the tilt and edge energies, i.e.,

    \begin{equation}
        \label{eq:cholesteric_surface_interaction}
        E' = \frac{C}{2}\int{[1-(\vu{u}\cdot\vu{n})^2]\dd{A}}+\lambda\oint{\dd{s}}.
    \end{equation}

    For a disk of area $A$ lying in the $xy$ plane, the perimeter is given by $\oint{\dd{s}}=2\sqrt{\pi A}$ and the tilt is given by

    \begin{equation}
        \begin{aligned}
            \int{[1-(\vu{u}\cdot\vu{n})^2]\dd{A}} & =A-\int_{0}^{(A/\pi)^{1/2}}\int_{0}^{2\pi}{\cos[2](q\rho \cos\phi)\rho\dd{\rho}\dd{\phi}} \\&=[1-J_1(2qR_0)/(qR_0)]A/2,
        \end{aligned}
    \end{equation}

    where $J_1(x)$ is the first order Bessel function of the first kind and $R_0=\sqrt{A/\pi}$ is the radius of the disk. Thus, we find

    \begin{equation}
        \label{eq:embedded_disk}
        E'_{disk} = 2\lambda\sqrt{\pi A}+\frac{C}{4}A[1-J_1(2qR_0)/(qR_0)].
    \end{equation}

    For Enneper's surface, the perimeter is

    \begin{equation}
        \label{eq:Enneper_perimeter}
        \oint{\dd{s}}=2\pi (1+r_1^{2m})R r_1,
    \end{equation}

    and the surface normal is given by

    \begin{equation}
        \label{eq:Enneper_normal}
        \vu{n}=\left(\frac{2r^m\cos(m\phi)}{1+r^{2m}},\frac{2r^m\sin(m\phi)}{1+r^{2m}},\frac{r^{2m}-1}{1+r^{2m}}\right).
    \end{equation}

    Writing the tilt in the $(r,\phi)$ coordinates,

    \begin{equation}
        \label{eq:Enneper_tilt}
        \int{[1-(\vu{u}\cdot\vu{n})^2]\dd{A}}=A-\int_{0}^{r_1}\int_{0}^{2\pi}{(\vu{u}\cdot\vu{n})^2r(1+r^{2m})^2R^2\dd{r}\dd{\phi} },
    \end{equation}

    we find

    \begin{equation}
        \begin{aligned}
            \label{eq:Enneper_energy}
            E'_{Enneper}  & = CA-C\int_{0}^{r_1}\int_{0}^{2\pi}{(\vu{u}\cdot\vu{n})^2r(1+r^{2m})^2R^2\dd{r}\dd{\phi}} \\
                                & + \lambda 2\pi (1+r_1^{2m})R r_1.
        \end{aligned}
    \end{equation}

    Although our simple model yields an analytic formula for the energy, it is not easy to analytically minimize the energy due to the lack of axisymmetry and its complicated dependence on $r_1$ through eqn (\ref{eq:Rr1}). Therefore, we minimize the energy over $r_1$ and $\vu{q}$ numerically using Powell's method.\cite{powell1964efficient} To avoid large numerical errors when calculating the tilt energy as $r_1\rightarrow 0$, we enforce $r_1>0.2$ in the minimization process. Likewise, we demand that $r_1<\sqrt{3}$ so that the ($m=1$) Enneper surface does not intersect itself.\cite{enneper}

    \begin{figure}[htb]
        \includegraphics[width=\linewidth]{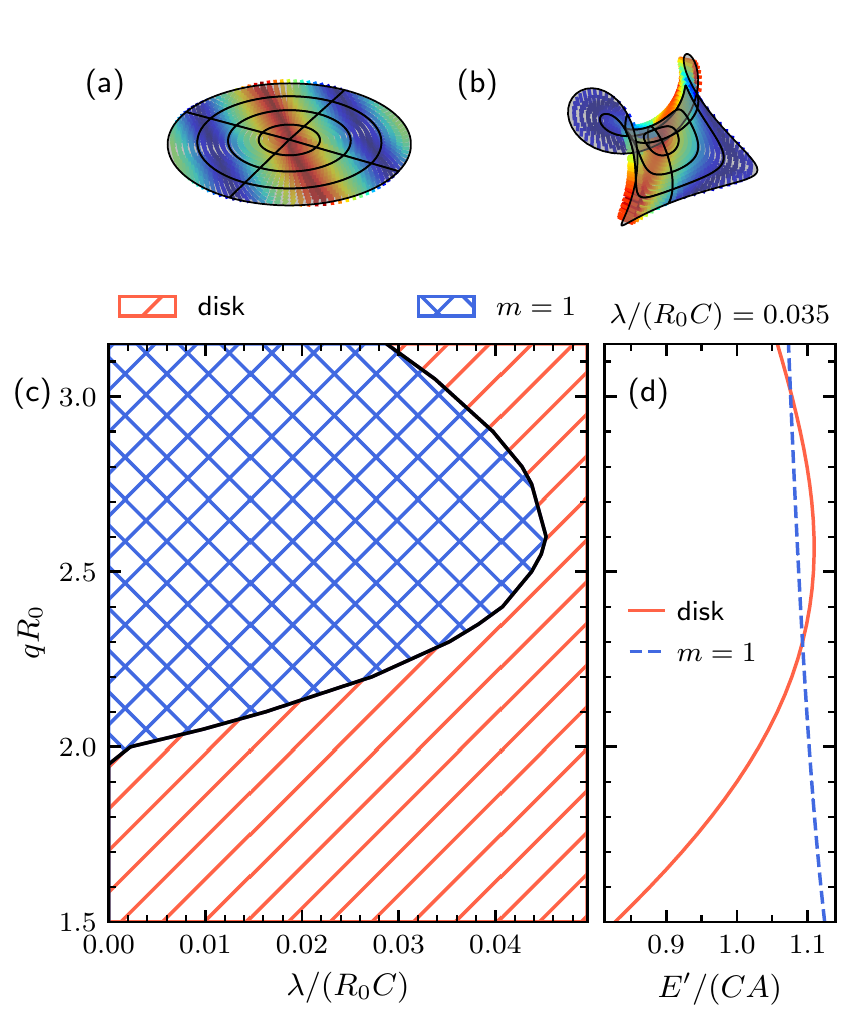}
        \caption{\label{fig:numerical_enneper} The transition from a disk to an Enneper surface in the simplified model.
        (a) A disk in the director field $\vu{u}$ of eqn (\ref{eq:3d_cholesteric}). The color indicates the tilt angle between the local surface normal and $\vu{u}$, as in Fig.~\ref{fig:config_demo}. The wire frame is a guide to the eye.
        (b) Enneper surface of order $m=1$ embedded embedded in the same director field as in (a). The shape is obtained by minimizing the energy with respect to $r_1$ and $\vu{q}$ with $qR_0=2.5$ and $\lambda/(R_0 C)=0.03$, yielding the optimized values $r_1=1.2$ and $\vu{q}\cdot\vu{x}=0.71$.
        (c) Shape phase diagram for the simplified model.
        (d) Energies of the optimal $m=1$ Enneper surface (dashed) and a disk (solid) vs. dimensionless wavenumber $qR_0$ for $\lambda/(R_0C)=0.035$.}
    \end{figure}

    By comparing the energies of the disk and the Enneper surface, we find the shape phase diagram shown in Fig.~\ref{fig:numerical_enneper}(a). Since a full turn of the directors in the cholesteric phase appears in a disk once the diameter of the disk exceeds the pitch, we only consider $qR_0>\pi/2$. Also, we see in Fig.~\ref{fig:negK_lam}(a) that when more than two $\pi$ walls are present, the walls are not parallel and cannot be described by eqn~(\ref{eq:3d_cholesteric}). Thus, to restrict our analysis to no more than two $\pi$ walls, we require that $qR_0 \leq \pi$. For sufficiently large $qR_0$ and small enough $\lambda$, Enneper's surface has lower energy than the flat disk. The critical value of $qR_0$ at which Enneper's surface is energetically favorable increases as $\lambda/(R_0 C)$ increases until reaching a critical point, beyond which a disk shape is always of lower energy. When $0.029\lesssim\lambda/(R_0 C)\lesssim0.044$, the state of minimum energy is a disk at small $qR_0$, then an $m=1$ Enneper's surface at larger $qR_0$, and then a disk again as $qR_0$ is further increased. This reentrance arises from the oscillation of the tilt energy with the cholesteric pitch, as represented by the term $J_1(2qR_0)/(qR_0)$ in $E'_{disk}$, eqn.~(\ref{eq:embedded_disk}). Fig.~\ref{fig:numerical_enneper}(b) illustrates this reentrance with the energies of the disk and Enneper surface at $\lambda/(R_0 C)=0.035$.

    The tilt configuration predicted by the simple model [Fig.~\ref{fig:numerical_enneper}(b)] is similar to the tilt configuration found by the Monte Carlo simulations [Fig.~\ref{fig:negK_lam}(a)]. In both cases, the rotation of the directors due to the cholesteric twist is the same as the rotation of the normals along a line that is at 45 degrees from the direction of steepest descent of the saddle. Although the normal vectors to a surface have no twist, the tilt interaction energy is lowered in the region where the directors are aligned with the normals.

    We also need to recognize the limitations of this simple analytical model. The single-twist director field eqn~(\ref{eq:3d_cholesteric}) is independent of the shape of the membrane surface. Thus, the twist of the director near the membrane edge is neglected, which would contribute to an effective edge bending energy that favors the disk shape. Nevertheless, this model still embodies the main idea that a cholesteric director field can drive the rippling of the membrane surface.

\section{Conclusion}
\label{sec:conclusion}
    In this paper we studied, using Monte Carlo simulations, a discrete model of chiral membranes composed of rod-like viruses. Our model allows us to consider the interplay of chirality, free edges and membrane shape with no \textit{a priori} assumptions of shape or director orientation. We found three phases of the orientational order: isotropic, smectic-A and cholesteric. In the smectic-A phase, the twist of directors is expelled to the membrane edge which is in agreement with experiment\cite{barry2009direct} and theory\cite{pelcovits2009twist}. The transition to a cholesteric phase leads to a rippling of the membrane with a saddle-like shape similar to what has been observed experimentally.\cite{sharma2020} Using an analytic model of a saddle (an Enneper surface of order one), we showed how this shape has lower energy than a flat disk for sufficiently large chirality and small edge line tension.

    Our model is general enough to allow future study of a myriad of remarkable shapes that have been observed experimentally,\cite{Khanra_etal2021} including higher-order saddles, catenoids and shapes with more openings. Many of these shapes occur when long and short viruses are mixed together. Natural generalizations of our model would be to include a Lebwohl-Lasher interaction with multiple values corresponding to the different pairs of species of rods, to add a nonzero Gaussian curvature modulus and to account for the depletion interaction.

\section*{Conflicts of interest}
    There are no conflicts to declare.

\section*{Acknowledgements}
    We thank Leroy Jia and Zvonimir Dogic for helpful discussions. This work was supported in part by the National Science Foundation through Grants No. MRSEC-1420382, CMMI-1634552 and CMMI-2020098.



\balance


\bibliography{rsc} 

\providecommand*{\mcitethebibliography}{\thebibliography}
\csname @ifundefined\endcsname{endmcitethebibliography}
{\let\endmcitethebibliography\endthebibliography}{}
\begin{mcitethebibliography}{48}
\providecommand*{\natexlab}[1]{#1}
\providecommand*{\mciteSetBstSublistMode}[1]{}
\providecommand*{\mciteSetBstMaxWidthForm}[2]{}
\providecommand*{\mciteBstWouldAddEndPuncttrue}
  {\def\EndOfBibitem{\unskip.}}
\providecommand*{\mciteBstWouldAddEndPunctfalse}
  {\let\EndOfBibitem\relax}
\providecommand*{\mciteSetBstMidEndSepPunct}[3]{}
\providecommand*{\mciteSetBstSublistLabelBeginEnd}[3]{}
\providecommand*{\EndOfBibitem}{}
\mciteSetBstSublistMode{f}
\mciteSetBstMaxWidthForm{subitem}
{(\emph{\alph{mcitesubitemcount}})}
\mciteSetBstSublistLabelBeginEnd{\mcitemaxwidthsubitemform\space}
{\relax}{\relax}

\bibitem[Amabilino(2009)]{amabilino2009chirality}
D.~B. Amabilino, \emph{Chirality at the nanoscale: nanoparticles, surfaces,
  materials and more}, John Wiley \& Sons, 2009\relax
\mciteBstWouldAddEndPuncttrue
\mciteSetBstMidEndSepPunct{\mcitedefaultmidpunct}
{\mcitedefaultendpunct}{\mcitedefaultseppunct}\relax
\EndOfBibitem
\bibitem[Bahr and Kitzerow(2001)]{bahr2001chirality}
C.~Bahr and H.-S. Kitzerow, \emph{Chirality in liquid crystals}, Springer,
  2001\relax
\mciteBstWouldAddEndPuncttrue
\mciteSetBstMidEndSepPunct{\mcitedefaultmidpunct}
{\mcitedefaultendpunct}{\mcitedefaultseppunct}\relax
\EndOfBibitem
\bibitem[Wagni{\`e}re(2007)]{wagniere2007chirality}
G.~H. Wagni{\`e}re, \emph{On chirality and the universal asymmetry: reflections
  on image and mirror image}, John Wiley \& Sons, 2007\relax
\mciteBstWouldAddEndPuncttrue
\mciteSetBstMidEndSepPunct{\mcitedefaultmidpunct}
{\mcitedefaultendpunct}{\mcitedefaultseppunct}\relax
\EndOfBibitem
\bibitem[Wright and Mermin(1989)]{wright1989crystalline}
D.~C. Wright and N.~D. Mermin, \emph{Reviews of Modern Physics}, 1989,
  \textbf{61}, 385\relax
\mciteBstWouldAddEndPuncttrue
\mciteSetBstMidEndSepPunct{\mcitedefaultmidpunct}
{\mcitedefaultendpunct}{\mcitedefaultseppunct}\relax
\EndOfBibitem
\bibitem[Seifert(1997)]{seifert1997configurations}
U.~Seifert, \emph{Advances in physics}, 1997, \textbf{46}, 13--137\relax
\mciteBstWouldAddEndPuncttrue
\mciteSetBstMidEndSepPunct{\mcitedefaultmidpunct}
{\mcitedefaultendpunct}{\mcitedefaultseppunct}\relax
\EndOfBibitem
\bibitem[Gibaud(2017)]{gibaud2017filamentous}
T.~Gibaud, \emph{Journal of Physics: Condensed Matter}, 2017, \textbf{29},
  493003\relax
\mciteBstWouldAddEndPuncttrue
\mciteSetBstMidEndSepPunct{\mcitedefaultmidpunct}
{\mcitedefaultendpunct}{\mcitedefaultseppunct}\relax
\EndOfBibitem
\bibitem[Kamien and Selinger(2001)]{KamienSelinger2001}
R.~D. Kamien and J.~V. Selinger, \emph{J. Phys. Condens. Matter}, 2001,
  \textbf{\textbf{13}}, R1\relax
\mciteBstWouldAddEndPuncttrue
\mciteSetBstMidEndSepPunct{\mcitedefaultmidpunct}
{\mcitedefaultendpunct}{\mcitedefaultseppunct}\relax
\EndOfBibitem
\bibitem[Oswald and Pieranski(2005)]{OswaldPieranski2005}
P.~Oswald and P.~Pieranski, \emph{Nematic and cholesteric liquid crystals},
  Taylor \& Francis, 2005\relax
\mciteBstWouldAddEndPuncttrue
\mciteSetBstMidEndSepPunct{\mcitedefaultmidpunct}
{\mcitedefaultendpunct}{\mcitedefaultseppunct}\relax
\EndOfBibitem
\bibitem[Duzgun \emph{et~al.}(2018)Duzgun, Selinger, and
  Saxena]{duzgun2018comparing}
A.~Duzgun, J.~V. Selinger and A.~Saxena, \emph{Physical Review E}, 2018,
  \textbf{97}, 062706\relax
\mciteBstWouldAddEndPuncttrue
\mciteSetBstMidEndSepPunct{\mcitedefaultmidpunct}
{\mcitedefaultendpunct}{\mcitedefaultseppunct}\relax
\EndOfBibitem
\bibitem[Gibaud \emph{et~al.}(2012)Gibaud, Barry, Zakhary, Henglin, Ward, Yang,
  Berciu, Oldenbourg, Hagan, Nicastro,\emph{et~al.}]{gibaud2012reconfigurable}
T.~Gibaud, E.~Barry, M.~J. Zakhary, M.~Henglin, A.~Ward, Y.~Yang, C.~Berciu,
  R.~Oldenbourg, M.~F. Hagan, D.~Nicastro \emph{et~al.}, \emph{Nature}, 2012,
  \textbf{481}, 348--351\relax
\mciteBstWouldAddEndPuncttrue
\mciteSetBstMidEndSepPunct{\mcitedefaultmidpunct}
{\mcitedefaultendpunct}{\mcitedefaultseppunct}\relax
\EndOfBibitem
\bibitem[Balchunas \emph{et~al.}(2020)Balchunas, Jia, Zakhary, Robaszewski,
  Gibaud, Dogic, Pelcovits, and Powers]{balchunas2020force}
A.~Balchunas, L.~L. Jia, M.~J. Zakhary, J.~Robaszewski, T.~Gibaud, Z.~Dogic,
  R.~A. Pelcovits and T.~R. Powers, \emph{Physical Review Letters}, 2020,
  \textbf{125}, 018002\relax
\mciteBstWouldAddEndPuncttrue
\mciteSetBstMidEndSepPunct{\mcitedefaultmidpunct}
{\mcitedefaultendpunct}{\mcitedefaultseppunct}\relax
\EndOfBibitem
\bibitem[Jia \emph{et~al.}(2017)Jia, Zakhary, Dogic, Pelcovits, and
  Powers]{jia2017chiral}
L.~L. Jia, M.~J. Zakhary, Z.~Dogic, R.~A. Pelcovits and T.~R. Powers,
  \emph{Physical Review E}, 2017, \textbf{95}, 060701\relax
\mciteBstWouldAddEndPuncttrue
\mciteSetBstMidEndSepPunct{\mcitedefaultmidpunct}
{\mcitedefaultendpunct}{\mcitedefaultseppunct}\relax
\EndOfBibitem
\bibitem[Sharma \emph{et~al.}()Sharma, Saikia, Khanra, and Dogic]{sharma2020}
P.~Sharma, L.~Saikia, A.~Khanra and Z.~Dogic, {unpublished}\relax
\mciteBstWouldAddEndPuncttrue
\mciteSetBstMidEndSepPunct{\mcitedefaultmidpunct}
{\mcitedefaultendpunct}{\mcitedefaultseppunct}\relax
\EndOfBibitem
\bibitem[Robaszewski \emph{et~al.}()Robaszewski, Jia, Powers, Pelcovits, and
  Dogic]{Robaszewski2020}
J.~Robaszewski, L.~Jia, T.~R. Powers, R.~A. Pelcovits and Z.~Dogic,
  {unpublished}\relax
\mciteBstWouldAddEndPuncttrue
\mciteSetBstMidEndSepPunct{\mcitedefaultmidpunct}
{\mcitedefaultendpunct}{\mcitedefaultseppunct}\relax
\EndOfBibitem
\bibitem[Tu and Ou-Yang(2003)]{tu2003lipid}
Z.~Tu and Z.~Ou-Yang, \emph{Physical Review E}, 2003, \textbf{68}, 061915\relax
\mciteBstWouldAddEndPuncttrue
\mciteSetBstMidEndSepPunct{\mcitedefaultmidpunct}
{\mcitedefaultendpunct}{\mcitedefaultseppunct}\relax
\EndOfBibitem
\bibitem[Kaplan \emph{et~al.}(2010)Kaplan, Tu, Pelcovits, and
  Meyer]{kaplan2010theory}
C.~N. Kaplan, H.~Tu, R.~A. Pelcovits and R.~B. Meyer, \emph{Physical Review E},
  2010, \textbf{82}, 021701\relax
\mciteBstWouldAddEndPuncttrue
\mciteSetBstMidEndSepPunct{\mcitedefaultmidpunct}
{\mcitedefaultendpunct}{\mcitedefaultseppunct}\relax
\EndOfBibitem
\bibitem[Tu and Pelcovits(2013)]{tu2013theory}
H.~Tu and R.~A. Pelcovits, \emph{Physical Review E}, 2013, \textbf{87},
  032504\relax
\mciteBstWouldAddEndPuncttrue
\mciteSetBstMidEndSepPunct{\mcitedefaultmidpunct}
{\mcitedefaultendpunct}{\mcitedefaultseppunct}\relax
\EndOfBibitem
\bibitem[Kang \emph{et~al.}(2016)Kang, Gibaud, Dogic, and
  Lubensky]{kang2016entropic}
L.~Kang, T.~Gibaud, Z.~Dogic and T.~Lubensky, \emph{Soft Matter}, 2016,
  \textbf{12}, 386--401\relax
\mciteBstWouldAddEndPuncttrue
\mciteSetBstMidEndSepPunct{\mcitedefaultmidpunct}
{\mcitedefaultendpunct}{\mcitedefaultseppunct}\relax
\EndOfBibitem
\bibitem[Gibaud \emph{et~al.}(2017)Gibaud, Kaplan, Sharma, Zakhary, Ward,
  Oldenbourg, Meyer, Kamien, Powers, and Dogic]{gibaud2017achiral}
T.~Gibaud, C.~N. Kaplan, P.~Sharma, M.~J. Zakhary, A.~Ward, R.~Oldenbourg,
  R.~B. Meyer, R.~D. Kamien, T.~R. Powers and Z.~Dogic, \emph{Proceedings of
  the National Academy of Sciences}, 2017, \textbf{114}, E3376--E3384\relax
\mciteBstWouldAddEndPuncttrue
\mciteSetBstMidEndSepPunct{\mcitedefaultmidpunct}
{\mcitedefaultendpunct}{\mcitedefaultseppunct}\relax
\EndOfBibitem
\bibitem[Yang \emph{et~al.}(2012)Yang, Barry, Dogic, and Hagan]{yang2012self}
Y.~Yang, E.~Barry, Z.~Dogic and M.~F. Hagan, \emph{Soft Matter}, 2012,
  \textbf{8}, 707--714\relax
\mciteBstWouldAddEndPuncttrue
\mciteSetBstMidEndSepPunct{\mcitedefaultmidpunct}
{\mcitedefaultendpunct}{\mcitedefaultseppunct}\relax
\EndOfBibitem
\bibitem[Xie \emph{et~al.}(2016)Xie, Pelcovits, and Hagan]{xie2016probing}
S.~Xie, R.~A. Pelcovits and M.~F. Hagan, \emph{Physical Review E}, 2016,
  \textbf{93}, 062608\relax
\mciteBstWouldAddEndPuncttrue
\mciteSetBstMidEndSepPunct{\mcitedefaultmidpunct}
{\mcitedefaultendpunct}{\mcitedefaultseppunct}\relax
\EndOfBibitem
\bibitem[Koibuchi(2008)]{koibuchi2008possible}
H.~Koibuchi, \emph{Physical Review E}, 2008, \textbf{77}, 021104\relax
\mciteBstWouldAddEndPuncttrue
\mciteSetBstMidEndSepPunct{\mcitedefaultmidpunct}
{\mcitedefaultendpunct}{\mcitedefaultseppunct}\relax
\EndOfBibitem
\bibitem[Ramakrishnan \emph{et~al.}(2010)Ramakrishnan, Kumar, and
  Ipsen]{ramakrishnan2010monte}
N.~Ramakrishnan, P.~S. Kumar and J.~H. Ipsen, \emph{Physical Review E}, 2010,
  \textbf{81}, 041922\relax
\mciteBstWouldAddEndPuncttrue
\mciteSetBstMidEndSepPunct{\mcitedefaultmidpunct}
{\mcitedefaultendpunct}{\mcitedefaultseppunct}\relax
\EndOfBibitem
\bibitem[Nguyen \emph{et~al.}(2013)Nguyen, Geng, Selinger, and
  Selinger]{nguyen2013nematic}
T.-S. Nguyen, J.~Geng, R.~L. Selinger and J.~V. Selinger, \emph{Soft Matter},
  2013, \textbf{9}, 8314--8326\relax
\mciteBstWouldAddEndPuncttrue
\mciteSetBstMidEndSepPunct{\mcitedefaultmidpunct}
{\mcitedefaultendpunct}{\mcitedefaultseppunct}\relax
\EndOfBibitem
\bibitem[Sreeja \emph{et~al.}(2015)Sreeja, Ipsen, and Kumar]{sreeja2015monte}
K.~Sreeja, J.~H. Ipsen and P.~S. Kumar, \emph{Journal of Physics: Condensed
  Matter}, 2015, \textbf{27}, 273104\relax
\mciteBstWouldAddEndPuncttrue
\mciteSetBstMidEndSepPunct{\mcitedefaultmidpunct}
{\mcitedefaultendpunct}{\mcitedefaultseppunct}\relax
\EndOfBibitem
\bibitem[Lubensky and MacKintosh(1993)]{LubenskyMacKintosh1993}
T.~C. Lubensky and F.~C. MacKintosh, \emph{Phys. Rev. Lett.}, 1993,
  \textbf{\textbf 71}, 1565\relax
\mciteBstWouldAddEndPuncttrue
\mciteSetBstMidEndSepPunct{\mcitedefaultmidpunct}
{\mcitedefaultendpunct}{\mcitedefaultseppunct}\relax
\EndOfBibitem
\bibitem[Selinger and Schnur(1993)]{SelingerSchnur1993}
J.~V. Selinger and J.~M. Schnur, \emph{Phys. Rev. Lett.}, 1993, \textbf{\textbf
  71}, 4091\relax
\mciteBstWouldAddEndPuncttrue
\mciteSetBstMidEndSepPunct{\mcitedefaultmidpunct}
{\mcitedefaultendpunct}{\mcitedefaultseppunct}\relax
\EndOfBibitem
\bibitem[Zakhary \emph{et~al.}(2014)Zakhary, Gibaud, Kaplan, Barry, Oldenbourg,
  Meyer, and Dogic]{piwalls}
M.~J. Zakhary, T.~Gibaud, C.~N. Kaplan, E.~Barry, R.~Oldenbourg, R.~B. Meyer
  and Z.~Dogic, \emph{Nature Communications}, 2014, \textbf{5}, 3063\relax
\mciteBstWouldAddEndPuncttrue
\mciteSetBstMidEndSepPunct{\mcitedefaultmidpunct}
{\mcitedefaultendpunct}{\mcitedefaultseppunct}\relax
\EndOfBibitem
\bibitem[Sharma \emph{et~al.}(2014)Sharma, Ward, Gibaud, Hagan, and
  Dogic]{rafts}
P.~Sharma, A.~Ward, T.~Gibaud, M.~F. Hagan and Z.~Dogic, \emph{Nature}, 2014,
  \textbf{513}, 77--80\relax
\mciteBstWouldAddEndPuncttrue
\mciteSetBstMidEndSepPunct{\mcitedefaultmidpunct}
{\mcitedefaultendpunct}{\mcitedefaultseppunct}\relax
\EndOfBibitem
\bibitem[Ding \emph{et~al.}(2020)Ding, Pelcovits, and Powers]{ding2020shapes}
L.~Ding, R.~A. Pelcovits and T.~R. Powers, \emph{Physical Review E}, 2020,
  \textbf{102}, 032608\relax
\mciteBstWouldAddEndPuncttrue
\mciteSetBstMidEndSepPunct{\mcitedefaultmidpunct}
{\mcitedefaultendpunct}{\mcitedefaultseppunct}\relax
\EndOfBibitem
\bibitem[Gompper and Kroll(1997)]{gompper1997network}
G.~Gompper and D.~M. Kroll, \emph{Journal of Physics: Condensed Matter}, 1997,
  \textbf{9}, 8795\relax
\mciteBstWouldAddEndPuncttrue
\mciteSetBstMidEndSepPunct{\mcitedefaultmidpunct}
{\mcitedefaultendpunct}{\mcitedefaultseppunct}\relax
\EndOfBibitem
\bibitem[Canham(1970)]{canham1970minimum}
P.~B. Canham, \emph{Journal of Theoretical Biology}, 1970, \textbf{26},
  61--81\relax
\mciteBstWouldAddEndPuncttrue
\mciteSetBstMidEndSepPunct{\mcitedefaultmidpunct}
{\mcitedefaultendpunct}{\mcitedefaultseppunct}\relax
\EndOfBibitem
\bibitem[Helfrich(1973)]{helfrich1973elastic}
W.~Helfrich, \emph{Zeitschrift f{\"u}r Naturforschung C}, 1973, \textbf{28},
  693--703\relax
\mciteBstWouldAddEndPuncttrue
\mciteSetBstMidEndSepPunct{\mcitedefaultmidpunct}
{\mcitedefaultendpunct}{\mcitedefaultseppunct}\relax
\EndOfBibitem
\bibitem[Espriu(1987)]{espriu1987triangulated}
D.~Espriu, \emph{Physics Letters B}, 1987, \textbf{194}, 271--276\relax
\mciteBstWouldAddEndPuncttrue
\mciteSetBstMidEndSepPunct{\mcitedefaultmidpunct}
{\mcitedefaultendpunct}{\mcitedefaultseppunct}\relax
\EndOfBibitem
\bibitem[Lebwohl and Lasher(1972)]{lebwohl1972nematic}
P.~A. Lebwohl and G.~Lasher, \emph{Physical Review A}, 1972, \textbf{6},
  426\relax
\mciteBstWouldAddEndPuncttrue
\mciteSetBstMidEndSepPunct{\mcitedefaultmidpunct}
{\mcitedefaultendpunct}{\mcitedefaultseppunct}\relax
\EndOfBibitem
\bibitem[Memmer(2000)]{memmer2000computer}
R.~Memmer, \emph{Liquid Crystals}, 2000, \textbf{27}, 533--546\relax
\mciteBstWouldAddEndPuncttrue
\mciteSetBstMidEndSepPunct{\mcitedefaultmidpunct}
{\mcitedefaultendpunct}{\mcitedefaultseppunct}\relax
\EndOfBibitem
\bibitem[Boal and Rao(1992)]{boal1992topology}
D.~H. Boal and M.~Rao, \emph{Physical Review A}, 1992, \textbf{46}, 3037\relax
\mciteBstWouldAddEndPuncttrue
\mciteSetBstMidEndSepPunct{\mcitedefaultmidpunct}
{\mcitedefaultendpunct}{\mcitedefaultseppunct}\relax
\EndOfBibitem
\bibitem[Barker and Watts(1969)]{barker1969structure}
J.~Barker and R.~Watts, \emph{Chemical Physics Letters}, 1969, \textbf{3},
  144--145\relax
\mciteBstWouldAddEndPuncttrue
\mciteSetBstMidEndSepPunct{\mcitedefaultmidpunct}
{\mcitedefaultendpunct}{\mcitedefaultseppunct}\relax
\EndOfBibitem
\bibitem[Sokal(1997)]{sokal1997monte}
A.~Sokal, \emph{Functional Integration}, Springer, 1997, pp. 142--145\relax
\mciteBstWouldAddEndPuncttrue
\mciteSetBstMidEndSepPunct{\mcitedefaultmidpunct}
{\mcitedefaultendpunct}{\mcitedefaultseppunct}\relax
\EndOfBibitem
\bibitem[Gompper and Kroll(2000)]{gompper2000melting}
G.~Gompper and D.~M. Kroll, \emph{The European Physical Journal E}, 2000,
  \textbf{1}, 153--157\relax
\mciteBstWouldAddEndPuncttrue
\mciteSetBstMidEndSepPunct{\mcitedefaultmidpunct}
{\mcitedefaultendpunct}{\mcitedefaultseppunct}\relax
\EndOfBibitem
\bibitem[de~Gennes(1972)]{de1972analogy}
P.~G. de~Gennes, \emph{Solid State Communications}, 1972, \textbf{10},
  753--756\relax
\mciteBstWouldAddEndPuncttrue
\mciteSetBstMidEndSepPunct{\mcitedefaultmidpunct}
{\mcitedefaultendpunct}{\mcitedefaultseppunct}\relax
\EndOfBibitem
\bibitem[Lin \emph{et~al.}(2015)Lin, Saxena, and Batista]{Lin2015}
S.-Z. Lin, A.~Saxena and C.~D. Batista, \emph{Phys. Rev. B}, 2015, \textbf{91},
  224407\relax
\mciteBstWouldAddEndPuncttrue
\mciteSetBstMidEndSepPunct{\mcitedefaultmidpunct}
{\mcitedefaultendpunct}{\mcitedefaultseppunct}\relax
\EndOfBibitem
\bibitem[Pelcovits and Meyer(2009)]{pelcovits2009twist}
R.~A. Pelcovits and R.~B. Meyer, \emph{Liquid Crystals}, 2009, \textbf{36},
  1157--1160\relax
\mciteBstWouldAddEndPuncttrue
\mciteSetBstMidEndSepPunct{\mcitedefaultmidpunct}
{\mcitedefaultendpunct}{\mcitedefaultseppunct}\relax
\EndOfBibitem
\bibitem[Barry \emph{et~al.}(2009)Barry, Dogic, Meyer, Pelcovits, and
  Oldenbourg]{barry2009direct}
E.~Barry, Z.~Dogic, R.~B. Meyer, R.~A. Pelcovits and R.~Oldenbourg, \emph{The
  Journal of Physical Chemistry B}, 2009, \textbf{113}, 3910--3913\relax
\mciteBstWouldAddEndPuncttrue
\mciteSetBstMidEndSepPunct{\mcitedefaultmidpunct}
{\mcitedefaultendpunct}{\mcitedefaultseppunct}\relax
\EndOfBibitem
\bibitem[Meyer \emph{et~al.}(2003)Meyer, Desbrun, Schr{\"o}der, and
  Barr]{meyer2003discrete}
M.~Meyer, M.~Desbrun, P.~Schr{\"o}der and A.~H. Barr, \emph{Visualization and
  Mathematics III}, Springer, 2003, pp. 35--57\relax
\mciteBstWouldAddEndPuncttrue
\mciteSetBstMidEndSepPunct{\mcitedefaultmidpunct}
{\mcitedefaultendpunct}{\mcitedefaultseppunct}\relax
\EndOfBibitem
\bibitem[Fomenko and Tuzhilin(1991)]{enneper}
A.~T. Fomenko and A.~A. Tuzhilin, \emph{Elements of the geometry and topology
  of minimal surfaces in three- dimensional space}, American Mathematical
  Society, 1991\relax
\mciteBstWouldAddEndPuncttrue
\mciteSetBstMidEndSepPunct{\mcitedefaultmidpunct}
{\mcitedefaultendpunct}{\mcitedefaultseppunct}\relax
\EndOfBibitem
\bibitem[Powell(1964)]{powell1964efficient}
M.~J. Powell, \emph{The Computer Journal}, 1964, \textbf{7}, 155--162\relax
\mciteBstWouldAddEndPuncttrue
\mciteSetBstMidEndSepPunct{\mcitedefaultmidpunct}
{\mcitedefaultendpunct}{\mcitedefaultseppunct}\relax
\EndOfBibitem
\bibitem[Khanra \emph{et~al.}(2021)Khanra, Jia, Mitchell, Balchunas, Pelcovits,
  Powers, Dogic, and Sharma]{Khanra_etal2021}
A.~Khanra, L.~L. Jia, N.~Mitchell, A.~Balchunas, R.~A. Pelcovits, T.~R. Powers,
  Z.~Dogic and P.~Sharma, \emph{unpublished}, 2021\relax
\mciteBstWouldAddEndPuncttrue
\mciteSetBstMidEndSepPunct{\mcitedefaultmidpunct}
{\mcitedefaultendpunct}{\mcitedefaultseppunct}\relax
\EndOfBibitem
\end{mcitethebibliography}
\bibliographystyle{rsc} 

\end{document}